\documentclass[aps,prl,reprint,superscriptaddress]{revtex4-1}

\usepackage{natbib}
\usepackage{amsmath} % AMS Math Package
\usepackage{amssymb}	% Math symbols such as \mathbb
\usepackage{graphicx} % Allows for eps images
\usepackage{times}
\usepackage{color}
\usepackage{bm}% bold math
\usepackage[section]{placeins}
\usepackage{setspace}
\usepackage{anysize}
\usepackage{graphicx}% Include figure files
\usepackage{epstopdf}
\usepackage[colorlinks=true,linkcolor=blue, citecolor=blue, urlcolor=blue, 
unicode=true]{hyperref}
\marginsize{2.0 cm}{2.0 cm}{1.5 cm}{1.5 cm}

\begin{document}
\newcommand\bbone{\ensuremath{\mathbbm{1}}}
\newcommand{\ul}{\underline}
\newcommand{\bp}{{\bf p}}
\newcommand{\vl}{v_{_L}}
\newcommand{\vc}{\mathbf}
\newcommand{\be}{\begin{equation}}
\newcommand{\ee}{\end{equation}}
\newcommand{\bk}{{{\bf{k}}}}
\newcommand{\bK}{{{\bf{K}}}}
\newcommand{\cE}{{{\cal E}}}
\newcommand{\bQ}{{{\bf{Q}}}}
\newcommand{\br}{{{\bf{r}}}}
\newcommand{\bg}{{{\bf{g}}}}
\newcommand{\bG}{{{\bf{G}}}}
\newcommand{\hbr}{{\hat{\bf{r}}}}
\newcommand{\bR}{{{\bf{R}}}}
\newcommand{\bq}{{\bf{q}}}
\newcommand{\hx}{{\hat{x}}}
\newcommand{\hy}{{\hat{y}}}
\newcommand{\hd}{{\hat{\delta}}}
\newcommand{\bea}{\begin{eqnarray}}
\newcommand{\eea}{\end{eqnarray}}
\newcommand{\ra}{\rangle}
\newcommand{\la}{\langle}
\renewcommand{\tt}{{\tilde{t}}}
\newcommand{\upa}{\uparrow}
\newcommand{\dna}{\downarrow}
\newcommand{\bS}{{\bf S}}
\newcommand{\vS}{\vec{S}}
\newcommand{\dg}{{\dagger}}
\newcommand{\pdg}{{\phantom\dagger}}
\newcommand{\tphi}{{\tilde\phi}}
\newcommand{\cf}{{\cal F}}
\newcommand{\ca}{{\cal A}}
\renewcommand{\ni}{\noindent}
\newcommand{\ct}{{\cal T}}
\newcommand{\brf}{\bar{F}}
\newcommand{\brg}{\bar{G}}
\newcommand{\jeff}{j_{\rm eff}}

\title{Evidence for negative thermal expansion in the superconducting precursor phase SmFeAsO}

\author{H.~D.~Zhou}
\affiliation{Department of Physics and Astronomy, University of Tennessee, Knoxville, TN 37996, USA}

\author{P.~M.~Sarte}
\affiliation{School of Chemistry, University of Edinburgh, Edinburgh EH9 3FJ, United Kingdom}
\affiliation{Centre for Science at Extreme Conditions, University of Edinburgh, Edinburgh EH9 3FD, United Kingdom}

\author{B.~S.~Conner}
\affiliation{Materials Science and Technology Division, Oak Ridge National Laboratory, Oak Ridge, TN 37831, USA}

\author{L.~Balicas}
\affiliation{National High Magnetic
	Field Laboratory, Florida State University, Tallahassee, FL
	32306-4005, USA}

\author{C.~R.~Wiebe}
\altaffiliation{Author to whom correspondence should be addressed: \href{mailto:ch.wiebe@uwinnipeg.ca}{ch.wiebe@uwinnipeg.ca}}
\affiliation{Department of Chemistry, University of Winnipeg, Winnipeg, MB R3B 2E9 Canada}
\affiliation{Department of Physics and Astronomy, McMaster University, Hamilton, ON L8S 4M1, Canada}
\affiliation{Canadian Institute for Advanced Research, Toronto, ON M5G 1Z8, Canada}
\affiliation{Department of Chemistry, University of Manitoba, Winnipeg, MB R3T 2N2, Canada}

\author{X.~H.~Chen}
\affiliation{Hefei National Laboratory for Physical Sciences at Microscale and Department of Physics, University of Science and Technology of China, Hefei, Anhui 230026, China}
\affiliation{Collaborative Innovation Center of Advanced Microstructures, Nanjing University, Nanjing 210093, China}
\affiliation{Key Laboratory of Strongly-coupled Quantum Matter Physics,
	University of Science and Technology of China, Chinese Academy of Sciences, Hefei 230026, China}
%\affiliation{High Magnetic Field Laboratory, Chinese Academy of Sciences, Hefei, Anhui 230031, China}

\author{T.~Wu}
\affiliation{Hefei National Laboratory for Physical Sciences at Microscale and Department of Physics, University of Science and Technology of China, Hefei, Anhui 230026, China}
\affiliation{Collaborative Innovation Center of Advanced Microstructures, Nanjing University, Nanjing 210093, China}
\affiliation{Key Laboratory of Strongly-coupled Quantum Matter Physics,
	University of Science and Technology of China, Chinese Academy of Sciences, Hefei 230026, China}

\author{G.~Wu}
\affiliation{Hefei National Laboratory for Physical Sciences at Microscale and Department of Physics, University of Science and Technology of China, Hefei, Anhui 230026, China}

\author{R.~H.~Liu}
%\affiliation{Hefei National Laboratory for Physical Sciences at Microscale and Department of Physics, University of Science and Technology of China, Hefei, Anhui 230026, China}
\affiliation{Department of Physics, Emory University, Atlanta, Georgia 30322, USA}

\author{H.~Chen}
\affiliation{Hefei National Laboratory for Physical Sciences at Microscale and Department of Physics, University of Science and Technology of China, Hefei, Anhui 230026, China}

\author{D.~F.~Fang}
\affiliation{Hefei National Laboratory for Physical Sciences at Microscale and Department of Physics, University of Science and Technology of China, Hefei, Anhui 230026, China}

\date{\today}% It is always \today, today, but any date may be explicitly specified

\begin{abstract}
	
	The fluorine-doped rare-earth iron oxypnictide series SmFeAsO$_{1-x}$F$_x$ (0 $\leq x \leq$ 0.10) was investigated with high resolution powder x-ray scattering. In agreement with previous studies [Margadonna \emph{et al.}, Phys. Rev. B. \textbf{79}(1), 014503 (2009)], the parent compound SmFeAsO exhibits a tetragonal-to-orthorhombic structural distortion at T$\rm{_{S}}$~=~130~K which is rapidly suppressed by $x \simeq$ 0.10 deep within the superconducting dome. The change in unit cell symmetry is followed by a previously unreported magnetoelastic distortion at 120~K. The temperature dependence of the thermal expansion coefficient $\alpha_{V}$ reveals a rich phase diagram for SmFeAsO: (i) a global minimum at 125 K corresponds to the opening of a spin-density wave instability as measured by pump-probe femtosecond spectroscopy [Mertelj \emph{et al.}, Phys. Rev. B \textbf{81}(22), 224504(2010)] whilst (ii) a global maximum at 110 K corresponds to magnetic ordering of the Sm and Fe sublattices as measured by magnetic x-ray scattering [Nandi \emph{et al.}, Phys. Rev. B. \textbf{84}(5), 055419 (2011)]. At much lower temperatures than T$\rm{_{N}}$, SmFeAsO exhibits a significant negative thermal expansion on the order of -40~ppm~$\cdot$~K$^{-1}$ in contrast to the behavior of other rare-earth oxypnictides such as PrFeAsO [Kimber \emph{et al.} Phys. Rev. B \textbf{78}(14), 140503 (2008)] and the actinide oxypnictide NpFeAsO [Klimczuk \emph{et al.} Phys. Rev. B \textbf{85}, 174506 (2012)] where the onset of $\alpha <$ 0 only appears in the vicinity of magnetic ordering.  Correlating this feature with the temperature and doping dependence of the resistivity and the unit cell parameters, we interpret the negative thermal expansion as being indicative of the possible condensation of itinerant electrons accompanying the opening of a SDW gap, consistent with transport measurements [Tropeano \emph{et al.}, Supercond. Sci. Technol. \textbf{22}(3), 034004 (2009)].      
	
%	The lattice distortions that are clearly visible with x-ray synchrotron measurements are detectable, although deep within the superconducting phase they appear only as orthorhombic strain due to the coarser resolution.  In the parent compound SmFeAsO, significant negative thermal expansion is observed below 50 K on the order of 35 ppm $\cdot$ K$^{-1}$.  Correlating this feature with a peak in the resistivity, we interpret this result as being due to the condensation of itinerant electrons.
\end{abstract}

\maketitle

\indent The discovery of iron oxypnictides of the general formula RFeAsO (R = RE$^{3+}$) has brought about a renaissance in the field of high temperature superconductivity~\cite{delacruz2008,stewart2011,liu2008}.  Previous efforts were almost completely focused on the cuprates for nearly two decades, with no real clear picture emerging for the superconducting mechanism or an explanation of the rich phase diagrams as a function of doping~\cite{aharony1988,vaknin1987,emery1987,bednorz1986,hussey2016,rybicki2016}.  It has been well-understood that strong spin-spin coupling, in addition to the two-dimensional layers of square planar CuO plaquettes, and mobility introduced through electron and/or hole doping of the Mott insulator phases are common properties to all of the superconducting phases~\cite{lee2006}.  However, the discovery of superconductivity in the two-dimensional iron oxypnictide stuctures (with the $d^{7}$ Fe$^{2+}$) has left an indelible mark on the condensed matter community~\cite{kamihara2008,chengf}.  A combination of the relatively high values of T$\rm{_{c}}$~\cite{singh2013,chen2008,zhi2008,delacruz2008,zhao2008,wang2008} for the earliest samples and the striking similarities of their rich phase diagrams to those of the cuprates~\cite{si2008,fang2008,mazin2008,xu2008,zhao2008} suggests that through further refinements of the chemistry, and an understanding of the mechanism, these iron oxypnictides and iron-based two-dimensional structures in general not only provide a potential route for advancing our understanding of the cuprates but may even provide T$\rm{_{c}}$ values challenging the records of the cuprates.\\
\indent In this paper, we focus on x-ray studies of the fluorine-doped oxypnictide series SmFeAsO$_{1-x}$F$_x$ - a series with one of the highest T$\rm{_c}$ values ($\sim$ 58~K~for optimal $x$~=~0.20~\cite{singh2013}) among the iron-based superconductors.  A prominent tetragonal-to-orthorhombic structural distortion is observed at T$\rm{_{S}}$~=~130 K in the parent compound SmFeAsO corresponding to a peak in $\frac{d\rho}{dT}$, quickly followed by a previously unreported magnetoelastic coupling at T$\rm{^{*}}$~=~120~K. The temperature dependence of the thermal expansion coefficient $\alpha_{V}$ reveals that the transition at 120~K lies between a global minimum of the thermal expansion corresponding to the opening of a spin-density wave (SDW) instability as measured by pump-probe femtosecond spectroscopy~\cite{mertelj2010} and a global maximum corresponding to magnetic ordering on both the Sm and Fe magnetic sublattices as measured by magnetic x-ray scattering~\cite{nandi2011}. Upon cooling to temperatures much below T$\rm{^{*}}$, SmFeAsO exhibits a significant negative thermal expansion, corresponding to a broad feature in $\frac{d\rho}{dT}$. Although previously unreported for SmFeAsO, this behavior is reminiscent of other iron oxypnictides such as PrFeAsO~\cite{kimber2008} and LaFeAsO~\cite{nomura2008}. Since this behavior has also been previously reported for LaFeAsO~\cite{nomura2008} with a non-magnetic RE$^{3+}$ site, the negative thermal expansion was interpreted as an effect due to electron localization, specifically upon the Fe site, a claim that was also alluded to by systematic transport measurements on SmFeAsO$_{1-x}$F$_{x}$~\cite{tropeano2009}. Our interpretation of the data is supported by the observed suppression of all anomalous features in the temperature dependence of the crystallographic parameters with minimal fluorine doping since it is well-documented that electron-doping across the RFeAsO$_{1-x}$F$_{x}$ (R = RE$^{3+}$) series induces superconductivity --- by $x \sim$ 0.07~\cite{ding2008,liu2008} for Sm --- and the onset of superconductivity quickly suppresses the tetragonal-to-orthorhombic distortion, the SDW phase and the N\'{e}el phase~\cite{margadonna2009,wu2013,dong2008,delacruz2008, zhao2008}. Although the negative thermal expansion is quickly suppressed with minimal fluorine doping, it is worthwhile to note that since the lowest temperature anomaly occurs at an energy scale which is close to T$\rm{_c}$ in optimally doped samples~\cite{zhi2008, singh2013,chen2008}, the negative thermal expansion may correspond to a SDW that is rapidly suppressed in the SmFeAsO$_{1-x}$F$_{x}$ series as superconductivity evolves.
%We find that the tetragonal-to-orthorhombic structural distortion persists at low levels of doping, but upon reaching the superconducting phase, the transition becomes weak, and to within the resolution of our instrument only shows an orthorhombic strain as opposed to a true phase transition.  Near optimal doping, the strain appears to be no longer present in the sample, retaining tetragonal symmetry from room temperature to at least 8~K.  This effect is reminiscent of the strain across the La$_{2-x}$Sr$_2$CuO$_4$ phase diagram as the superconducting phase is approached with hole doping. All anomalous 

Polycrystalline samples of SmFeAsO$_{1-x}$F$_x$ ($x$ = 0, 0.05 and 0.10) were prepared by a two-step solid state reaction as previously outlined by Tropeano \emph{et al.}~\cite{tropeano2009}. The precursor SmAs was obtained by reacting high purity Sm and As metal at 600$\rm{^{o}}$C for 3~h, and then at 900$\rm{^{o}}$C for 5~h.  A mixture of SmAs, Fe, Fe$_2$O$_3$, and FeF$_2$ powders with the appropriate nominal stoichiometric ratio was then ground thoroughly and pressed into small pellets.  These pellets were wrapped in Ta foil and sealed in an evacuated quartz tube and annealed at 1200$\rm{^{o}}$C for 24~h. The pellets were then retrieved, reground, repressed and annealed at 1300$\rm{^{o}}$C for a further 72~h.  Resistivity measurements were performed by the conventional four-point-probe method using a Quantum Design PPMS.  The powder x-ray diffraction (pXRD) patterns were recorded by a HUBER Imaging Plate Guinier Camera 670 with monochromatized Cu $K\rm{_{\alpha,1}}$ radiation. Polycrystalline samples of SmFeAsO$_{1-x}$F$_{x}$ were cooled to a base temperature of 10~K with a closed cycle cryostat and held at base temperature for 12 hours, whilst after each temperature change, the samples were held at the desired temperature for 10 minutes with the objective of achieving thermal equilibrium and avoiding spurious features such as phase trapping. 
%The value of the thermal expansion coefficient $\alpha_{V}$ was calculated numerically with both \texttt{MATLAB R2016a} and \texttt{OriginPro v9.4.} using the forward difference method with a $\delta T$ of 5~K for the parent compound and 10~K for the fluorine doped members.   

%pXRD patterns down to 10 K were obtained. 

\begin{figure}[h]
	\centering
	\includegraphics[width=1.0\linewidth]{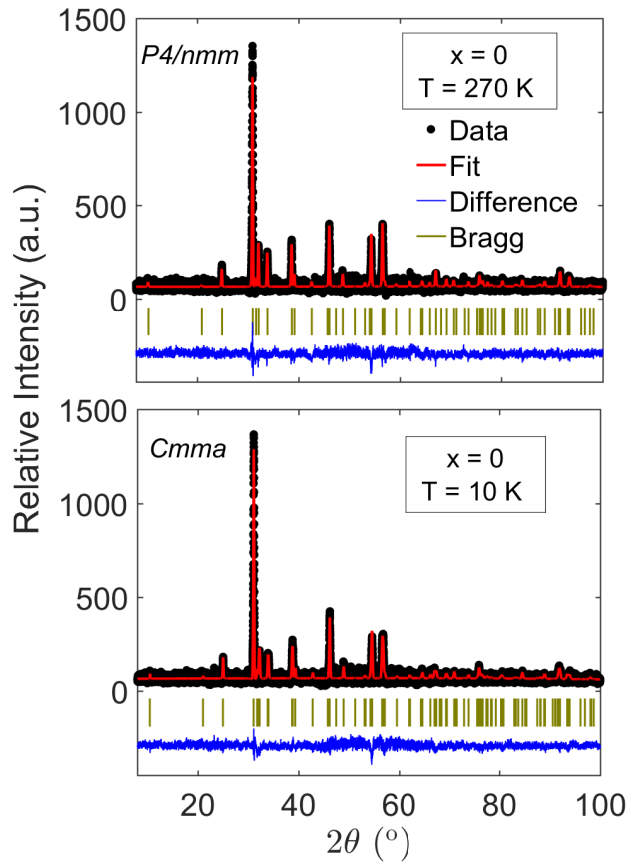}
	\caption{Measured, calculated and difference high resolution laboratory powder x-ray powder diffraction profiles for SmFeAsO at 270~K (tetragonal) and 10~K (orthorhombic) measured on a HUBER Imaging Plate Guinier Camera 670 with Cu K$_{\alpha,1}$ radiation. The Bragg reflections' locations of the tetragonal \emph{P4/nmm} (270~K) and orthorhombic \emph{Cmma} (10~K) phases are given by the olive vertical lines.}
	\label{fig:figure1}
\end{figure}

\begin{figure}
	\centering
	\includegraphics[width=1.0\linewidth]{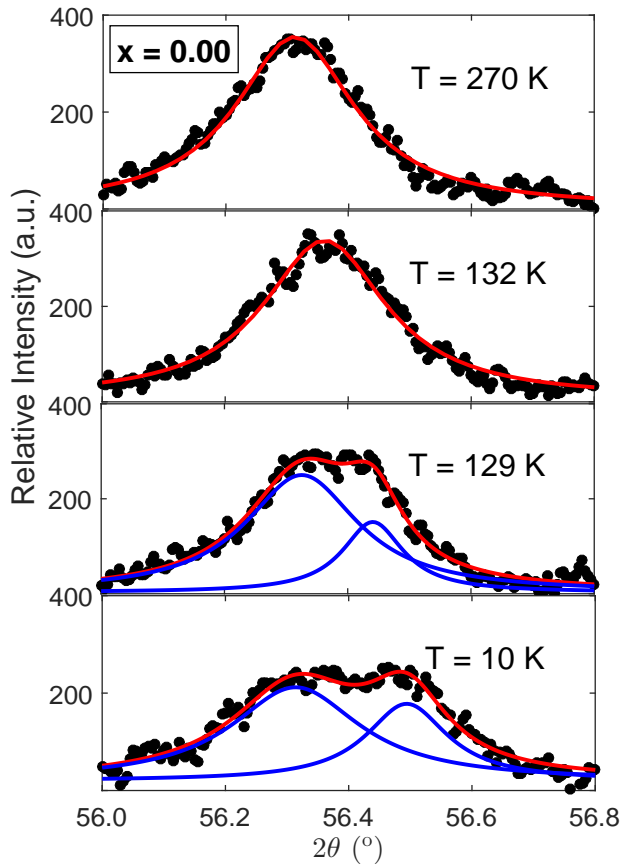}
	\caption{Temperature dependence of the (212) Bragg peak in the parent compound SmAsFeO ($x$~=~0).  The tetragonal-to-orthorhombic structural distortion at T$\rm{_{S}}$~=~130~K corresponds to the splitting of the (212) peak in the high temperature tetragonal \emph{P4/nmm} to the (132) and (312) peaks in the low temperature orthorhombic \emph{Cmma} phase. Solid red curves are the sum of fits to (132) and (312) individual peaks (solid blue curves) for the \emph{Cmma} phase or the (212) peak in the high temperature \emph{P4/nmm} phase.}
	\label{fig:figure2}
\end{figure}

\begin{figure}
	\centering
	\includegraphics[width=1.05\linewidth]{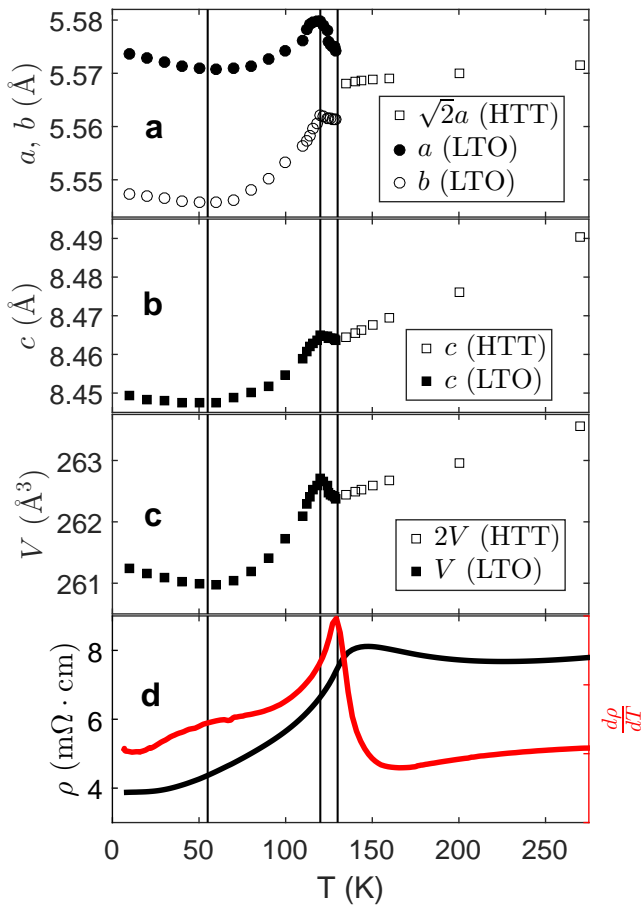}
	\caption{Temperature dependence of the $a$, $b$, and $c$ lattice parameters for SmAsFeO (panels (a) and (b)) revealing a first order transition accompanying the tetragonal-to-orthorhombic distortion noted in Fig.~\ref{fig:figure2}. A strong magnetoelastic response at 120~K attributed to magnetic ordering (T$\rm{_{N}}$~=~110~K) following the opening of the SDW phase (T$\rm{_{SDW}}$~=~125~K) and is followed by a prominent negative thermal expansion in all three crystallographic phases beneath 55 K, far from the vincinity of any magnetic ordering. (c) The temperature dependence of the unit cell volume, exhibiting an anomaly at 120 K, followed by negative thermal expansion beneath 55 K.  (d) The temperature dependence of the resistivity, and its derivative $\frac{d\rho}{dT}$.  Note that a peak in $\frac{d\rho}{dT}$ occurs at 130 K and a broad anomaly appears as well at 55~K, coinciding with the onset of the tetragonal-to-orthorhombic distortion and the onset of negative thermal expansion, respectively. The labels \texttt{HTT} and \texttt{LTO} denote \textbf{h}igh \textbf{t}emperature \textbf{t}etragonal phase and \textbf{l}ow \textbf{t}emperature \textbf{o}rthorhombic phase, respectively. \textbf{N.B.} Error bars are smaller than the size of the symbol representation of the experimental data.}
	\label{fig:figure3}
\end{figure}

\begin{figure}
	\centering
		\hspace*{-10.00mm}
	\includegraphics[width=1.01\linewidth]{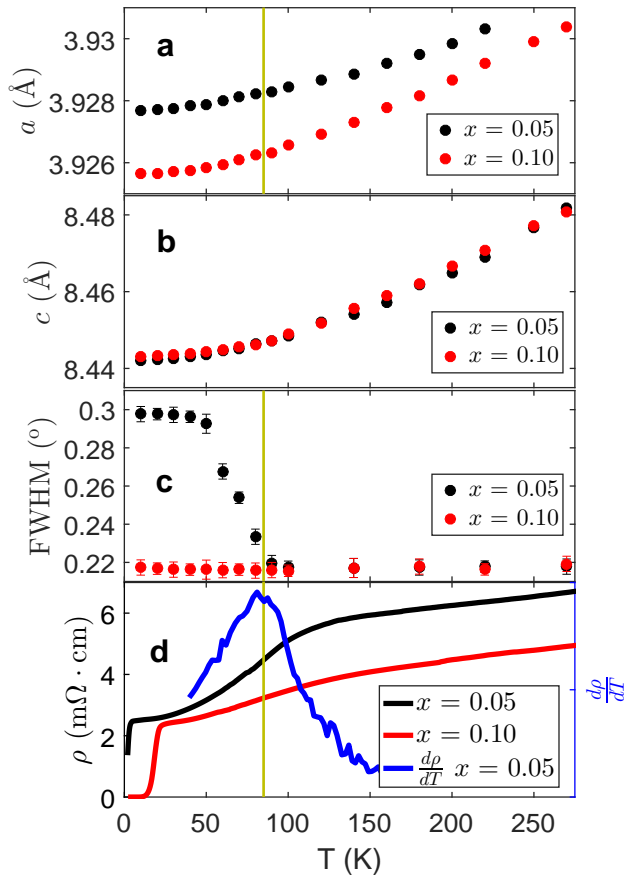}
	\caption{Comparison of the temperature dependence of the (a) $a$ and (b) $c$ lattice parameters for fluorine-doped members of the iron oxypnicitide series SmFeAsO$_{1-x}$F$_{x}$ ($x$~=~0.05, 0.1). All anomalous features in Fig.~\ref{fig:figure3} appear to have been completely suppressed with fluorine doping, consistent with the onset of superconductivity~\cite{delacruz2008}. (c) Temperature dependence of the FHWM of the (212) peak identified a distinct broadening for $x$~=~0.05, indicating the tetragonal-to-orthorhombic distortion persists with small amounts of fluorine doping, supported by the observation of a small kink in the temperature dependence of $a$ in panel (a), consistent with previous synchrotron results~\cite{margadonna2009}. The distortion appears to be completely suppressed by $x$~=~0.10.  (d) Temperature dependence of resistivity  reveals that the high temperature upturn is substantially suppressed by $x$~=~0.05 and completely suppressed by $x$~=~0.10. A distinct peak in $\frac{d\rho}{dT}$ is well-correlated with both the onset of the broadening of the (212) peak and the kink in the temperature dependence of $a$ for the $x$~=~0.05 sample indicated by the olive vertical line.}
	\label{fig:figure4}
\end{figure}

Figure~\ref{fig:figure1} shows the Rietveld refinement of the parent compound SmAsFeO ($x$~=~0). The pXRD data was fit using the \texttt{FULLPROF/WINPLOTR} suite~\cite{rodriguez1993,roisnel1999}.  As summarized by Tab.~\ref{tab:1}, the refined lattice parameters and atomic positions correlate well with previous structural studies confirming that SmFeAsO adopts the layered ZrCuSiAs-type structure corresponding to the tetragonal \emph{P4/nmm} space group~\cite{margadonna2009,martinelli2009,quebe2000}. A prominent structural distortion at T$\rm{{_S}}$~=~130~K was deduced by the observation of splitting of structural Bragg peaks. One such example includes the splitting of the (212) to the (312) and (132) as illustrated in Fig.~\ref{fig:figure2}, corresponding to a structural transition from the tetragonal \emph{P4/nmm} to orthorhombic \emph{Cmma}, once again in agreement with previous synchrotron and physical property results~\cite{margadonna2009,martinelli2009, johrendt2011, ding2008,tropeano2009}. As summarized in Fig.~\ref{fig:figure3}$(d)$, the resistivity of the parent compound shows a broad anomaly at higher temperatures but below this anomaly, there are no obvious features or phase transitions, consistent with previous transport measurements~\cite{liu2008,kaciulis2010,tropeano2009}. By taking the derivative of the resistivity, we find that there is a clear maximum at 130~K corresponding to the tetragonal-to-orthorhombic distortion temperature T$\rm{_{S}}$. As shown in Fig.~\ref{fig:figure4}, the tetragonal-to-orthorhombic structural transition is quickly suppressed within the superconducting phase, in agreement with previous synchrotron results that deduced the complete suppression of the structural distortion by $x$~$\sim$~0.14~\cite{margadonna2009}.  Although the structural transition for $x$~=~0.05 was detected through both the broadening of the (212) Bragg peak and minor kink in the temperature evolution of $a$ at 80~K which is well correlated with a peak in $\frac{d\rho}{dT}$ as illustrated in Figs.~\ref{fig:figure4}$(a)$,$(c)$ and $(d)$, there is no such indication for the highest doped sample ($x$~=~0.10) due to the coarser resolution of the laboratory x-ray diffractometer used in this current study compared to ID31 at the ESRF utilized in previous studies~\cite{margadonna2009}. 

%\begin{figure}[tbp]
%\linespread{1}
%\par
%\includegraphics[width=90mm,height=75mm,angle=0]{figure18.eps}
%\par
%\caption{XRD patterns for SmAsFeO at
%(a) 270 K and (b) 10 K.  The solid curves are the best fits from the
%Rietveld refinement using FullProf.  The vertical marks indicate the
%position of Bragg peaks, and the bottom curves show the difference
%between the observed and calculated intensities.}
%\end{figure}

\begin{table}
		\caption{Refined crystallographic parameters at 270~K for SmAsFeO  from the Rietveld refinement\footnote{Space group, \emph{P4/nmm}.  At 270~K, $a$~=~3.9402(2)~{\AA}, $c$~=~8.4803(4)~{\AA}, $V$ = 263.56(6) \AA$^{3}$, R$_{p}$~$\approx$~6.0\%, R$_{wp}$~$\approx$~6.0\% and $\chi^2$ $\approx$ 3.01.} of the high resolution laboratory x-ray powder-diffraction profile. Numbers in parentheses indicate statistical errors.} \vspace*{2mm}
\begin{tabular}{|c|c|c|c|c|c|}
	\hline 
	Atom & Wyckoff Site  & $x$  & $y$  & $z$  & B$\rm{_{iso}}$ ({\AA}$^2$)  \\ 
	\hline 
Sm	& 2$c$ & 0.25 & 0.25  & 0.1364(3) & 1.13(4)    \\ 
	\hline 
Fe	& 2$b$  & 0.75 & 0.25  &  0.5 & 1.57(7)  \\ 
	\hline 
As	& 2$c$ & 0.25  & 0.25  & 0.6632(5)   & 1.20(3)  \\ 
	\hline 
O	& 2$a$ & 0.75 & 0.25  & 0 & 1.90(5)  \\ 
	\hline 
\end{tabular}

\label{tab:1} 
\end{table}

%$a$~=~3.9402(2)~{\AA} and  $c$~=~8.4803(4)~{\AA}.}

From Rietveld refinements, the temperature evolution of the lattice parameters was calculated for the parent phase and are presented in Figs.~\ref{fig:figure3}$(a)$-$(c)$ below. The first-order nature of the jump in both $a$ and $b$ is clear, in contrast with a more smooth evolution of $c$. Below T$\rm{_{S}}$, a significant structural distortion was observed at 120~K. This previously unreported distortion manifests itself as a decrease of $a$, $b$, and $c$, and therefore a cusp-like feature of the unit cell volume $V$.  A likely candidate for the 120~K transition is magnetic ordering accompanying a spin-density wave instability of the nested Fermi surface that is reminiscent of other rare-earth iron oxypnictides~\cite{delacruz2008,mazin2008,yin2008, dong2008}. Such a claim is supported by numerous observations in literature including the detection of spin-density wave ordering by pump-probe femtosecond spectroscopy at T$\rm{_{SDW}}\simeq$~125~K~\cite{mertelj2010} and the detection of long-range ordering of the Sm and Fe magnetic sublattices by magnetic x-ray scattering at T$\rm{_{N}}\simeq$~110~K~\cite{nandi2011}. The formation of a SDW state below T$\rm{_{S}}$ is further supported by broad features in both DC susceptometry and heat capacity reported in literature~\cite{nandi2011,meena2013}. It is worthwhile to note that there exists a slight increase in the resistivity of the parent compound (Fig.~\ref{fig:figure3}) as one decreases temperature before both T$\rm{_{S}}$ and T$\rm{_{SDW}}$ in agreement with previous transport measurements~\cite{liu2008,kaciulis2010,tropeano2009}. This behavior of the resistivity may correspond to the onset of magnetic correlations, reminisent of ``stripe" phase of La$_{1.6-x}$Nd$_{0.4}$Sr$_{x}$CuO$_{4}$~\cite{ichikawa2000}. Furthermore, multiple studies~\cite{klauss2008,mcguire2008,kimber2008} have correlated a broad feature in the resistivity at high temperatures proceeding T$\rm{_{S}}$ with the formation of SDW phase. As illustrated in Fig.~\ref{fig:figure4}$(d)$, this increase in the resistivity is quickly suppressed with fluorine-doping $x\simeq$~0.10, corresponding to the onset of superconductivity in SmFeAsO$_{1-x}$F$_{x}$ ($x$~$\sim$~0.07~\cite{ding2008,liu2008}) and as is the case for other high temperature superconductors, the concurrent destruction of the SDW state and its accompanying magnetic ordering transition~\cite{liu2008,kamihara2008}. The interpretation of the 120~K feature as magnetic ordering is strongly supported by the temperature dependence of the thermal expansion coefficient $\alpha_{V}$~=~$\frac{1}{V}\frac{\partial V}{\partial T}$ as presented in Fig.~\ref{fig:figure5} below. The temperature dependence of $\alpha_{V}$ reveals a rich phase diagram where a global minimum and maximum corresponds to the aforementioned literature reported values of T$\rm{_{SDW}}$~\cite{mertelj2010} and T$\rm{_{N}}$~\cite{nandi2011}, respectively; whilst, the magnetoelastic transition at 120~K corresponds to the crossover between negative and positive thermal expansion, indicating that the 120~K transition corresponds to some exotic phase of SDW and N\'{e}el phase coexistence, both phases that ultimately compete~\cite{delacruz2008,rotter2009} with superconductivity and would be expected --- and as is observed --- to be quickly suppressed with fluorine doping.        \\
%is unclear with a lack of neutron scattering experiments.  Specific heat measurement show a broad anomaly existing between 120 K and 130 K which is completely quenched by x = 0.05 doping levels.  The resistivity of the parent compound shows a broad anomaly at higher temperatures (at 145 K, see the final panel of figure 3), but below this temperature there are no obvious features or phase transitions.  Taking the derivative of the resistivity, we find that there is an anomaly at 130 K which agrees with the structural phase transition temperature.  However, there is no prominent signature of a phase transition at 120 K in transport property measurements.  Future experiments on single crystals are needed to confirm this, and also to resolve the broad feature seen in thermodynamic probes such as the heat capacity.
\indent As shown in Figs.~\ref{fig:figure3} and~\ref{fig:figure5}, as SmFeAsO is cooled much below T$\rm{_{SDW}}$, another anomalous feature appears at T~$\approx$~55~K --- a negative, nearly isotropic thermal expansion exists in all three crystallographic directions. The presence of a negative thermal expansion for SmFeAsO, although not been previously reported in either high resolution synchrotron~\cite{margadonna2009} or three terminal capacitance measurements~\cite{klingeler2010}, has been detected in other rare-earth~\cite{kimber2008} and actinide oxypnictides~\cite{klimczuk2012}. The distinguishing feature of SmFeAsO is that the onset of negative thermal expansion does not coincide with any particular magnetic ordering process such as in NpFeAsO where the onset of $\alpha<$~0 coincides with T$\rm{_{N}}$~\cite{klimczuk2012}. To gain some insight on the microscopic origin of the negative thermal expansion, we shall compare this system to other known materials that show negative thermal expansion. For a recent review on negative thermal expansion, please refer to Chen \emph{et al.}~\cite{chen2015}. One obvious quantitative parameter for comparison is the coefficient thermal expansion $\alpha_{V}$~=~$-40$ ppm $\cdot$ K$^{-1}$ at 10~K for the undoped compound. Simple flourite structures with tetrahedrally coordinated atoms experience small negative thermal expansion with $\alpha_{V}$ $\gtrsim$ $-10$~ppm~$\cdot$~K$^{-1}$~\cite{white1978l,barrera2005,lightfoot2001,quartz,attfield2016}. The tilting of rigid polyhedra in oxides with $O$-$M$-$O$ bridging in materials such as ZrW$_2$O$_8$, heralded as a compound with significant negative thermal expansion, yields a $\alpha_{V}$ of $-27.3$~ppm~$\cdot$~K$^{-1}$~\cite{mary1996,sleight1998}. Since SmAsFeO has a  significantly larger value, it is unlikely that such a mechanism exists. \\ 
%\begin{figure}[tbp]
%\linespread{1}
%\par
%\includegraphics[width=110mm,height=90mm,angle=0]{figure24.eps}
%\par
%\caption{Temperature dependence of the (212) Bragg peak in SmAsFeO (x=0).  The low temperature structural transition to orthorhombic symmetry is clear with the splitting of the (212) peak to the (132) and (312) peaks within the revised unit cell.}
%\end{figure}
%\begin{figure}[tbp]
%\linespread{1}
%\par
%\includegraphics[width=90mm,height=120mm,angle=0]{figure39.eps}
%\par
%\caption{Temperature dependence of the a, b, and c axis lattice parameters for SmAsFeO (x=0 phase, panels (a) and (b)).  Note the negative thermal expansion in all three crystallographic phases beneath 50 K. (c)  The temperature dependence of the net volume, showing an anomaly at 120 K, followed by negative thermal expansion beneath 50 K.  (d) The temperature dependence of the resistivity, and the derivative.  Note that a peak in the derivative occurs at 130 K (at the structural transition), and a broad anomaly appears as well at 50 K.HTT: high temperature tetragonal phase; LTO: low temperature orthorhombic phase.}
%\end{figure}
\indent A possible alternative explanation for the detection of negative thermal expansion could be attributed to the presence of Sm. The $f$ block elements have a rich and complex series of structural phase diagrams as a function of temperature and pressure~\cite{taylor1972,kotliar2001,soderlind1990}. The iron oxypnictides, with the addition magnetic Fe sublattice~\cite{nandi2011,ryan2009}, further complicates the effects of magneto-elastic coupling and $f$-electron physics in these materials. Among the rare earth compounds, crystal field induced negative thermal expansions have been noted on the order of -2~ppm~$\cdot$~K$^{-1}$ in compounds such as TmTe~\cite{ott1977,bergmann2013}. However, other prominent effects are noted in the vicinity of magnetic transitions, such as a coefficient of -500~ppm~$\cdot$~K$^{-1}$ near the Curie temperature in holmium~\cite{white1989}.  However, in the absence of features in the susceptibility, it is unlikely that this is the origin of the effect in SmAsFeO.  The one example that bears the greatest similarity to SmAsFeO is the change in the electronic configuration of Sm in Sm$_{2.72}$C$_{60}$~\cite{boucherle2013,arvanitidis2003}. In this compound, a truly dramatic change in the negative thermal expansion is observed below 50~K and is believed to be due to the change in size of the Sm ion of the 4$f^6d^0$ and 4$f^5d^1$ electronic configurations~\cite{grima2010} with  very similar transitions seen throughout other rare-earth systems below 60~K~\cite{mattens1980,yb,kuznetsov2003}.  The possibility of such an electronic transition in SmFeAsO can be quickly discredited by noting refinements indicate the average size increase of Sm is less than 1\% and observing there exists a small but detectable negative thermal expansion below 50~K for LaFeAsO~\cite{nomura2008}. Combining the observation of $\alpha_{V}<$~0 for LaFeAsO and the assumption that the underlying mechanism in LaFeAsO is similar to that in SmFeAsO, provides an argument against Sm electron localization, since the La$^{3+}$ ions should not adopt a valence fluctuating state.

%\begin{figure}[tbp]
%\linespread{1}
%\par
%\includegraphics[width=90mm,height=80mm,angle=0]{figure4correct1.eps}
%\par
%\caption{Comparison of the a and c lattice parameters for the x = 0.05 and x = 0.1 phases.  (c)  FHWM of the (212) peak for the x = 0.05 phase, indicating a broadening due to the lattice distortion at 100 K.  (d)  The resistivity and derivative for the x = 0.05 and x = 0.1 phases.}
%\end{figure}

Ruling out the presence of rare-earth valence fluctuations, one can now turn to the condensation of electrons as a possible mechanism.  The iron oxypnictides are itinerant electron systems with a reduced ordered iron moment consistently below $\sim$ 0.8~$\mu_{B}$ within the SDW regime~\cite{delacruz2008,kimber2008,zhao2008,chen2008magnetic}. The condensation of electrons from a higher occupied band to a lower band, for example, could be a possible mechanism for the negative thermal expansion, as is seen in other itinerant metals such as Cr (-~9~ppm~$\cdot$~K$^{-1}$)~\cite{grima2010,takenaka2012}, a metal that also possesses a nested Fermi surface~\cite{fawcett1988}. In fact, current models of itinerant magnetism in the iron pnictides have been successful in predicting the \textbf{Q}-wavevector of the incommensurate ordering, and have provided an explanation for the structural phase transition as a function of doping, albeit there are key differences between various models at the present that are highly dependent on sensitive parameters~\cite{mazin2008}. Furthermore, resistivity, magnetoresistivity, Hall effect, Seebeck coefficient, infrared reflectivity measurements performed by Tropeano \emph{et al.}~\cite{tropeano2009} on SmFeAsO$_{1-x}$F$_{x}$ ($x$~=~0 and 0.07) alluded to a condensation of electrons from the opening of a SDW gap. In agreement with previous measurements~\cite{liu2008,kaciulis2010,tropeano2009}, we note that there is a signature for the condensation of electrons in the resistivity. As shown in Fig.~\ref{fig:figure3}$(d)$, the derivative of the resistivity exhibits a broad peak at approximately 55~K, coinciding with the onset of the negative thermal expansion and thus supporting our interpretation of the negative thermal expansion as a consequence of electronic condensation. The energy scale, on the order of T$\rm{_c}$ in optimally doped samples~\cite{zhi2008, singh2013,chen2008}, suggests the possibility that the proposed electron condensation may play a key role in the mechanism of superconductivity of SmFeAsO$_{1-x}$F$_{x}$.  The suppression of the negative thermal expansion down to 10~K for the fluorine-doped samples as illustrated in Figs.~\ref{fig:figure4} and~\ref{fig:figure5}, suggests that the negative thermal expansion cannot be solely attributed to the superconducting phase but instead to the competing SDW phase; a phase that is also rapidly suppressed with the onset of superconductivity~\cite{wu2013,dong2008,delacruz2008}. Finally, it is worthwhile to note that $\mu$SR measurements~\cite{carlo2009} on a variety of underdoped iron-arsenic superconductors such as LaFeAsO$_{1-x}$F$_x$ ($x$~=~0.03) have revealed the presence of a Bessel function line shape to the relaxation, reminicent of the behavior of the cuprates within the stripe-ordered phase~\cite{savici2002}.  This line-shape is distinct from the parent compound LaOFeAs, and suggests that there is an electronic condensation leading to a reduced field at the muon site. Consequently, the negative thermal expansion noted in this current work may possibly be an indication of this condensation associated with stripe-like order which would be consistent with the magnetic ordering on the Sm and Fe sublattices~\cite{nandi2011}. \\
\begin{acknowledgments}	
	We acknowledge useful conversations with J.P.~Attfield, A.J.~Browne, K.H.~Hong, G.M.~McNally and S.E.~Maytham. This work utilized facilities supported in part by the NSF under Agreement No.~DMR-0084173. A portion of this work was made possible by the NHMFL In-House Research Program, the Schuller Program, the EIEG Program, the State of Florida, and the DOE.  C.R.W. would like to specifically acknowledge the Canada Research Chair (Tier II) program, CIFAR, NSERC, CFI and the ACS Petroleum Fund for financial support. P.M.S. acknowledges funding from the CCSF and the University of Edinburgh through the GRS and PCDS. L.B. acknowledges financial support from the DOE-BES Program through Award No.~DE-SC0002613.	
\end{acknowledgments}

% and opens the field to future searches for possible stripe ordering in the pnictides.

%For the sample with minimal doping ($x$~=~0.05) below the critical doping of $x \sim$ 0.07, there is a broad peak in the resistivity at elevated temperatures (100 K) which is likely due to spin density wave ordering.  However, as previously this feature is severely suppressed in the superconducting phase at $x$ = 0.10, as is expected for the SDW transition in high T$\rm{_{c}}$'s.  We therefore ascribe the strong magnetoelastic coupling and negative thermal expansion as having an association with the magnetic phase transition and not the superconducting phase transition.

\begin{figure}[h!]
	\centering
	\includegraphics[width=1.0\linewidth]{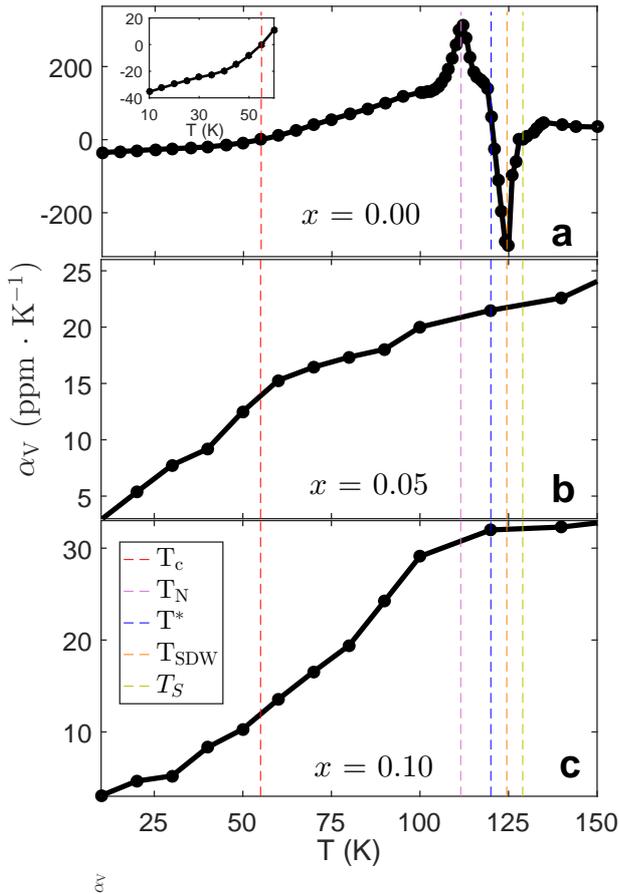}
	\caption{Temperature dependence of the thermal expansion coefficient $\alpha_{V}$ for (a) $x$~=~0.00, (b) $x$~=~0.05 and (c) $x$~=~0.10 members of the iron oxypnicitide series SmFeAsO$_{1-x}$F$_{x}$. For the parent compound $x$~=~0, the global maximum at 110~K and global minimum at 125~K correspond to magnetic ordering on the Sm and Fe magnetic sublattices as determined by magnetic x-ray scattering~\cite{nandi2011} and onset of the SDW phase as determined by pump-probe femtosecond spectroscopy~\cite{mertelj2010}, respectively. The three temperatures at which the value of $\alpha_{V}$ changes sign corresponds to the tetragonal-to-orthorhombic distortion temperature T$\rm{_{S}}$, the magneto-elastic distortion temperature T$^{*}$ and the critical temperature for optimally doped samples T$\rm{_{C}}$~\cite{zhi2008, singh2013,chen2008}. As illustrated in the inset of (a), the parent compound exhibits a significant negative thermal expansion $\alpha_{V}$~of~$-40$~ppm~$\cdot$~K$^{-1}$ below T$\rm{_{c}}$. The concurrent suppression of the negative thermal expansion at low temperatures and the anamolous features associated with the opening of the SDW gap with the fluroine-doped samples as illustrated in (b) and (c) strongly supports the interpretation of the negative thermal expansion as electron condensation due to the opening of the SDW gap as proposed by Tropeano \emph{et al.}~\cite{tropeano2009} deduced from transport measurements.}	
	\label{fig:figure5}
\end{figure}

%In summary, high resolution x-ray diffraction from 300 K to 8 K on the fluorine-doped iron oxypnictide series SmFeAsO$_{1-x}$F$_{x}$ has revealed three key features at 130 K, 120 K and 55 K. The first transition corresponds to the well-established tetragonal-to-orthorhombic structural distortion that is common among the rare-earth iron oxypnictides. The second transition most likely corresponds to ordering on the Sm and Fe magnetic sublattices accompanying the onset of a SDW phase. Whilst the final transition, corresponding to negative thermal expansion previously unreported in the literature, may be attributed to the localization of electrons on the Fe sites. The quick repression of the first two excitations within the superconducting regime suggests that these transitions are associated with magnetic order and not the superconducting transition, whilst the proximity of the third transition to the T$\rm{_{c}}$ of optimally doped samples suggests the localization of electrons may have a significant role in the superconductivity of not only SmFeAsO$_{1-x}$F$_{x}$ but Fe-based superconductors as a whole.       

%\bibliographystyle{nature}
\bibliography{SFAO_F_PMS_JPCM}

%\bibliographystyle{nature}
%\bibliography{Supplementary_Material}
%\bibliography{references_Co}
\end{document}